\documentclass{article}
\usepackage{spconf,amsmath,graphicx}

\usepackage{float}
\usepackage{numprint}
\usepackage{subfigure}

\npdecimalsign{.}
\newcommand{\eer}[1]{\nprounddigits{2}\numprint{#1}}
\newcommand{\dcf}[1]{\nprounddigits{3}\numprint{#1}}

\title{STUDYING SQUEEZE-AND-EXCITATION USED IN CNN FOR SPEAKER VERIFICATION}
\name{Mickael Rouvier, Pierre-Michel Bousquet}
\address{LIA, Avignon University}

\begin{document}

\maketitle
%


\begin{abstract}
In speaker verification, the extraction of voice representations is mainly based on the Residual Neural Network (ResNet) architecture. ResNet is built upon convolution layers which learn filters to capture local spatial patterns along all the input, then generate feature maps that jointly encode the spatial and channel information. Unfortunately, all feature maps in a convolution layer are learnt independently (the convolution layer does not exploit the dependencies between feature maps) and locally. This problem has first been tackled in image processing. A channel attention mechanism, called {\it squeeze-and-excitation} (SE), has recently been proposed in convolution layers and applied to speaker verification. This mechanism re-weights the information extracted across features maps. In this paper, we first propose an original qualitative study about the influence and the role of the SE mechanism applied to the speaker verification task at different stages of the ResNet, and then evaluate several SE architectures. We finally propose to improve the SE approach with a new pooling variant based on the concatenation of mean- and standard-deviation-pooling. Results showed that applying SE only on the first stages of the ResNet allows to better capture speaker information for the verification task, and that significant discrimination gains on Voxceleb1-E, Voxceleb1-H and SITW evaluation tasks have been noted using the proposed pooling variant.
\end{abstract}


\section{Introduction}
\label{sec:intro}

Speaker recognition refers to the task of verifying the identity claimed by a speaker from that person's voice~\cite{bimbot2004tutorial}. For example, it has been shown useful for speaker diarization~\cite{rouvier2012global}, forensics~\cite{campbell2009forensic} or voice dubbing~\cite{gresse2017acoustic}.

In recent years, Deep Neural Networks (DNN) have allowed to propose original voice representations, outperforming the state-of-the-art $i$-vector framework~\cite{dehak2010front}. One of this DNN approach seeks to extract an embedding representation of a speaker directly from its acoustic excerpts. This high-level speaker representation is called $x$-vector~\cite{snyder2018x}. In the $x$-vector framework, the DNN uses a stack of convolution layers followed by a temporal pooling layer that computes the mean and standard deviation of an input sequence. 

Convolutional layers allow us to extract hierarchical information from the speech signal in feature maps. Lower layers find trivial information such as low-level speech information, and upper layers capture more complex speech information. Unfortunately, whether high or low layers, extraction of feature maps by convolutional layers is carried out independently and locally (except for the last high-level layer).

In~\cite{hu2018squeeze}, the authors propose to tackle this problem by creating a new architectural unit which allows us to re-weight each feature map in convolutional layers by explicitly modelling inter-dependencies between feature maps. This mechanism performs feature recalibration by using global information in order to select useful information and suppress useless one. This method is called {\it squeeze-and-excitation} (SE).

The use of SE in Convolutional Neural Networks (CNN) for speaker verification has recently be introduced in~\cite{desplanques2020ecapa,brummerbut+,lee2020nec}; but its influence in details has never been studied on this speaker task. In this paper, we propose to study and improve the squeeze-and-excitation mechanism. In details, our contributions are as follows:

\begin{itemize}

    \item studying the influence of SE at different stages in ResNet-34 architecture. We observe that the best system is the one where SE is integrated on the first two stages of ResNet;
    
    \item evaluating different configurations of SE in the context of speaker verification, related to the pooling layer, integration strategy... We show that the parameters and architectures of SE that obtain the best performance in image processing are not the same than in speaker verification;
    
    \item obtaining a new variant of SE in which the global information generated by the SE by using the concatenation of two poolings: mean and standard-deviation;
    
\end{itemize}

Our experiments on the Speaker In The Wild (SITW) and Voxceleb1-E corpus without SE obtain respectively 1.39\% and 1.26\% of Equal Error Rate (EER), whereas the best system using SE obtains respectively 1.29\% and 1.13\% EER. A relative gain of 9\% is observed in terms of EER.

The paper is organized as follows: Section~\ref{sec:xvector} summarizes the $x$-vector approach.  Section~\ref{sec:squeeze} presents the squeeze-and-excitation (SE) approach. In Section~\ref{sec:expe_results}, we analyze the results of SE on speaker verification task and the contribution of our proposals. A qualitative study of the role of squeeze-and-excitation is proposed in Section~\ref{sec:study}. A conclusion is finally provided in Section~\ref{sec:conclusion}.

\section{$x$-vector system}
\label{sec:xvector}

An $x$-vector is a high-level speaker feature extracted from a DNN model. The DNN model is trained through a speaker identification task, {\it i.e.} by classifying speech segments into one of $n$ speaker identities. In that context, the different layers of the DNN are trained to extract information for discriminating between different speakers. The idea is to use one of the hidden layer as the speaker representation (the $x$-vector). One of the main advantage is that $x$-vectors produced by the DNN generalize well to speakers beyond those present in the training set. The benefits of $x$-vectors in terms of speaker detection accuracy have been demonstrated during the recent evaluation campaigns on NIST SRE~\cite{villalba2018jhu,lee2019nec,rouvier2019}, VoxCeleb 2020~\cite{thienpondt2020idlab,brummerbut+,torgashov2020id}, SdSVC~\cite{villalba_jhu_2019,bousquet_sdsvc_2019,barbadillo_2019}...

The $x$-vector extractor proposed in this paper is a variant based on ResNet~\cite{zeinali2019but}. The detailed topology is shown in Table~\ref{tbl:resnet34}.

\begin{table}[h]
    \caption{\label{tbl:resnet34}The proposed ResNet34 architecture. Last row, $N$ is the number of speakers. Batch-norm and ReLU layers are not shown. The dimensions are (Frequency$\times$Channels$\times$Time). The input is comprised of 60 filter banks from speech segments. A fixed segment length of 400 is used during training.}    
    \resizebox{\columnwidth}{!}{
    \begin{tabular}{l c c}
        \hline
        \hline
        \textbf{Layer name}   & \textbf{Structure}          & \textbf{Output} \\
        \hline
        Input                 & --                          & 60 $\times$ 400 $\times$ 1  \\
        Conv2D-1              & 3 $\times$ 3, Stride 1      & 60 $\times$ 400 $\times$ 128 \\
        \hline
        ResNetBlock-1         & $\begin{bmatrix} 3 \times 3, 128  \\ 3 \times 3, 128  \end{bmatrix} \times 3$  , Stride 1& $60\times 400 \times128$  \\
        ResNetBlock-2         & $\begin{bmatrix} 3 \times 3, 128  \\ 3 \times 3, 128  \end{bmatrix} \times 4$, Stride $2$ & $30 \times 200 \times 128$  \\
        ResNetBlock-3         & $\begin{bmatrix} 3 \times 3, 128 \\ 3 \times 3, 256 \end{bmatrix} \times 6$, Stride $2$ & $15 \times 100  \times 256$ \\
        ResNetBlock-4         & $\begin{bmatrix} 3 \times 3, 256 \\ 3 \times 3, 256 \end{bmatrix} \times 3$, Stride $2$ & $8  \times 50  \times 256$ \\
        \hline
        Pooling          & --                & $8 \times 256$                 \\
        Flatten               & --                & $2048$                            \\
        \hline
        Dense1                & --                & $256$                            \\
        Dense2 (Softmax)      & --                & $N$                              \\
        \hline
        Total                 & --                & --                             \\
        \hline
        \hline
    \end{tabular}
    }
\end{table}

\section{Squeeze-and-excitation}
\label{sec:squeeze}

As mentioned in Section~\ref{sec:intro}, the problem of convolution layers is that features maps are extracted independently and locally. In~\cite{hu2018squeeze}, the authors proposed to tackle this problem by creating a new architectural unit called squeeze-and-excitation (SE) block. The SE block allows us to model inter-dependencies of feature maps, so that the network is able to increase its sensitivity to the most informative features.

\begin{figure}[h]
\begin{center}
\includegraphics[width=0.17\textwidth]{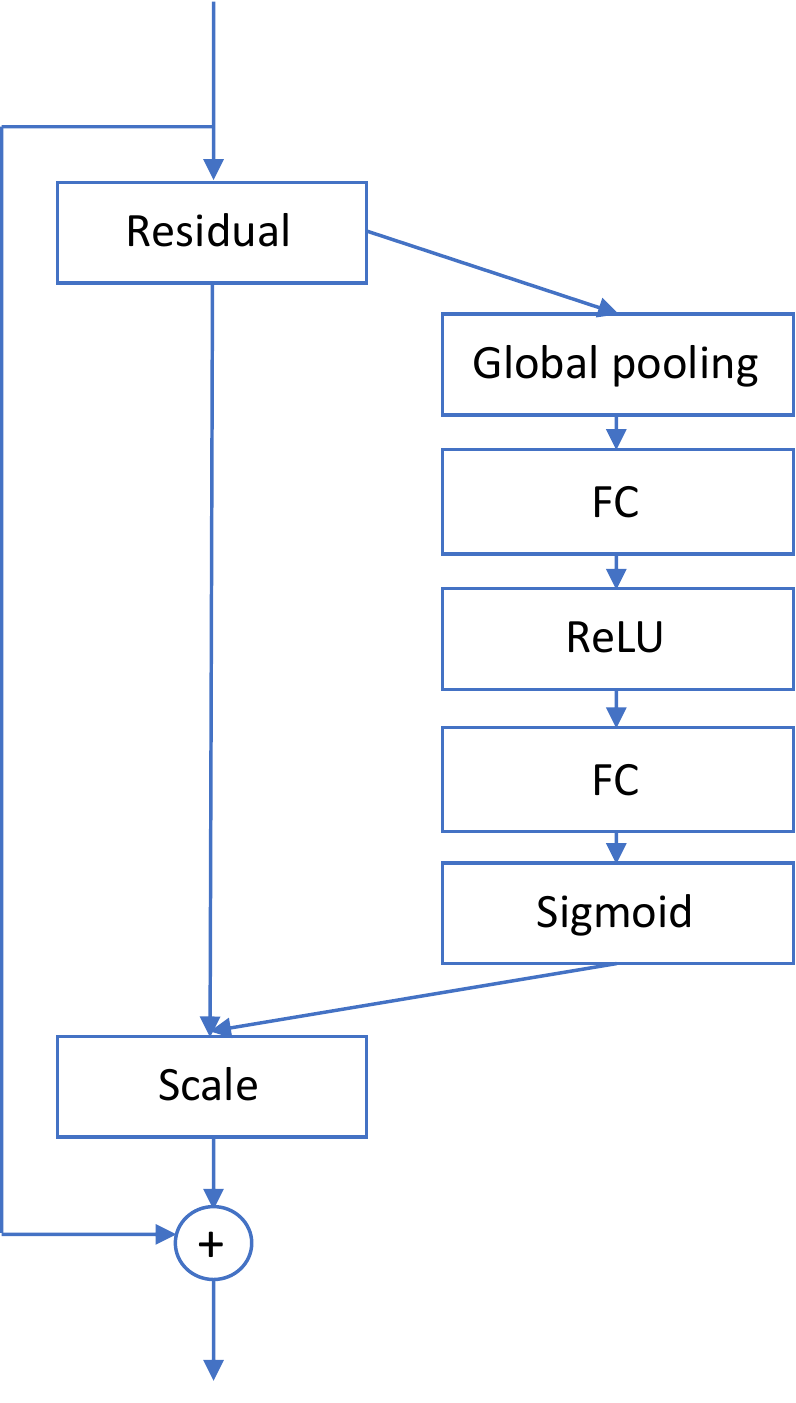}
\caption{Structure of the SE Block used in ResNet architecture.}
\label{fig:seblocks}
\end{center}
\end{figure}

The structure of the SE block is depicted in Figure~\ref{fig:seblocks}. First, a pooling layer is used to produce a global information of each channel by aggregating feature maps across their spatial dimension to a single numeric value. Thus, a vector of size $n$ is obtained where $n$ is equal to the number of feature maps. Afterwards, the vector is introduced into a two-layer neural network. A $n$ dimensional output vector is then obtained. These $n$ values can now be used as weights on the original feature maps, scaling each channel based on its importance. The SE process is performed in two steps: 1) produce a global information ({\it squeeze} step); and 2) re-weight each feature maps ({\it excitation} step).

The global information is achieved using a pooling layer. The pooling layer plays a central role in the SE strategy. While the mean-pooling obtains the best performance in image processing, it is unclear which pooling strategy performs best on speaker verification task. We then propose to evaluate different pooling strategies on speaker verification such as max-pooling, standard-deviation-pooling and the concatenation of mean- and standard-deviation pooling. 

Also, in order to limit model complexity, the hidden layers in SE blocks can be used as a reduction block where the input space is reduced to a smaller space and then expanded to the original dimensionality as the input. A discussion on this reduction is done in the experiments as well as the number of hidden layers.

The SE block can be simply integrated in CNN by inserting after the non-linearity following each convolution. In the case of ResNet, the classical integration strategy is to insert SE Block after the final convolutional layer and before the skip connection branch. The idea to integrate the SE Block before the skip connection branch, is to avoid noise in the skip connection branch and facilitate the learning of identity.

\section{Experiments and protocols}
\label{sec:expe_results}

This section describes the experimental setup in terms of dataset and experimental protocols.

\subsection{Training and Evaluation datasets}

The $x$-vector extractors are trained on the VoxCeleb2 dataset~\cite{chung2018voxceleb2}, only on the development partition, which contains speech excepts from 5,994 speakers with a 16 Khz sampling rate. The trained $x$-vectors are assessed on the Speakers in the Wild (SITW) core-core task~\cite{mclaren2016speakers}, Voxceleb1-E Cleaned and Voxceleb1-H Cleaned~\cite{nagrani2017voxceleb} dataset with a 16 KHz sampling rate. Note that the development set of VoxCeleb2 is completely disjoint from the VoxCeleb1 dataset ({\it i.e.} no common speakers).

We report results in terms of Equal Error Rate (EER) and the minimum of the normalized detection cost function (minDCF) at PTarget = $10^{-2}$.

\subsection{Implementation details}

The $x$-vector extractor used in this paper is a variant based on ResNet-34. The extractor cuts training dataset into 4-second chunks and augmented with noise, as described in~\cite{snyder2018x} and available as a part of the Kaldi-recipe. As input, we used 60-dimensional filter-banks. The $x$-vectors are 256-dimensional and the loss is the angular additive margin with scale equals to 30 and margin equals to 0.4. The size of the feature maps are 128, 128, 256 and 256 for the 4 ResNet blocks. We use stochastic gradient descent with momentum equals to 0.9, a weight decay equals to 2.10$^{-4}$ and initial learning rate equals to 0.2. The batch size was set to 128, however, training on 4 GPUs in parallel. The implementation is based on PyTorch and the model training takes about 2 days. In order to remove silence and low energy speech segments, a simple energy-based VAD is used based on the C0 component of the acoustic feature.

Let us note that, for a fair comparison, mini-batch used during neural network training and weights initialisation of neural networks are the same for all the experiments.

\subsection{SE blocks at different stages}

Table~\ref{tbl:stages} explores the influence by integrating SE blocks at different stages into the ResNet-34 (one stage at a time). The system called \emph{Baseline} is the system without SE blocks. This system obtained on Voxceleb1-E Cleaned and SITW 1.26\% and 1.39\% EER respectively. The system that achieves the best performance is the one that integrated SE blocks at Stage 1 and 2 (system called \emph{Stage=1,2}). This system obtained on Voxceleb1-E Cleaned and SITW 1.14\% and 1.99\% EER respectively. We observe that integrating SE blocks at all stages, as it is done in image processing, obtained the worst results  (1.30\% and 2.21\% EER respectively on Voxceleb1-E Cleaned and SITW).

\begin{table}[H]
\center
\caption{Results obtained by integrating SE Blocks at different stages.}
\label{tbl:stages}
\resizebox{\columnwidth}{!}{
\begin{tabular}[|c]{|l|l|c|c|c|c|c|c|}
\hline
\textbf{System}  & \multicolumn{2}{c|}{\textbf{VoxCeleb1}} & \multicolumn{2}{c|}{\textbf{VoxCeleb1}} & \multicolumn{2}{c|}{\textbf{SITW}}\\
& \multicolumn{2}{c|}{\textit{-E cleaned}} & \multicolumn{2}{c|}{\textit{-H cleaned}} & \multicolumn{2}{c|}{\textit{core-core}}\\
 & EER & DCF & EER & DCF & EER & DCF \\\hline
Baseline  & \eer{1.261} & \dcf{0.1312} &  \eer{2.12} & \dcf{0.1995} & \eer{1.394} &  \dcf{0.1269}  \\
Stage=1  & \eer{1.199} & \dcf{0.1290} &  \eer{2.039} & \dcf{0.1906} & \eer{1.34} &  \dcf{0.1147}  \\
Stage=1,2 & \textbf{\eer{1.142}} & \textbf{\dcf{0.1271}} &  \textbf{\eer{1.985}} & \textbf{\dcf{0.1869}} & \eer{1.312} &  \textbf{\dcf{0.1105}}  \\
Stage=1,2,3 & \eer{1.216} & \dcf{0.1331} &  \eer{2.05} & \dcf{0.1950} & \textbf{\eer{1.258}} &  \dcf{0.1160}  \\
Stage=1,2,3,4 & \eer{1.303} & \dcf{0.1336} &  \eer{2.21} & \dcf{0.2170} & \eer{1.394} &  \dcf{0.1300}  \\
\hline
\end{tabular}
}
\end{table}

\subsection{Reduction factor}

The SE Blocks is composed of two fully-connected hidden layers. These hidden layers can be used as a reduction block where the input space is reduced to a smaller space defined by the reduction factor ($r$) and then expanded to the original dimensionality as the input. Table~\ref{tbl:reduction_factor} investigates the trade-off between performance and model complexity by varying this reduction factor. We observe that  performance is robust for a reduction factor between $2$ and $4$. In image processing, the reduction factor is classically set to $16$. We observe that the system without any reduction factor obtained the best performance, but setting the reduction factor set to $4$ achieves a good balance between accuracy and complexity.

\begin{table}[H]
\center
\caption{Results obtained by using different reduction and expansion rates.}
\label{tbl:reduction_factor}
\resizebox{\columnwidth}{!}{
\begin{tabular}[|c]{|l|l|c|c|c|c|c|c|}
\hline
\textbf{System}  & \multicolumn{2}{c|}{\textbf{VoxCeleb1}} & \multicolumn{2}{c|}{\textbf{VoxCeleb1}} & \multicolumn{2}{c|}{\textbf{SITW}}\\
& \multicolumn{2}{c|}{\textit{-E cleaned}} & \multicolumn{2}{c|}{\textit{-H cleaned}} & \multicolumn{2}{c|}{\textit{core-core}}\\
 & EER & DCF & EER & DCF & EER & DCF \\\hline
Baseline & \textbf{\eer{1.142}} & \dcf{0.1271} &  \eer{1.985} & \dcf{0.1869} & \eer{1.312} &  \dcf{0.1105}  \\
r=2 & \eer{1.172} & \dcf{0.1271} &  \eer{1.998} & \dcf{0.1875} & \textbf{\eer{1.258}} &  \dcf{0.1149}  \\
r=4  & \eer{1.151} & \textbf{\dcf{0.1215}} &  \textbf{\eer{1.945}} & \textbf{\dcf{0.1807}} & \eer{1.34} &  \textbf{\dcf{0.1087}}  \\
r=8  & \eer{1.218} & \dcf{0.1358} &  \eer{2.09} & \dcf{0.1928} & \eer{1.558} &  \dcf{0.1178}  \\
\hline
\end{tabular}
}
\end{table}

\subsection{Integration strategy}

Table~\ref{tbl:integration} studies the influence of the location of the SE block when integrating into ResNet-34. In addition to the standard integration, three variants are proposed, similar to the ones proposed in~\cite{hu2018squeeze} and depicted in Figure~\ref{fig:integration} :

\begin{figure}[H]
\begin{center}
\includegraphics[width=0.45\textwidth]{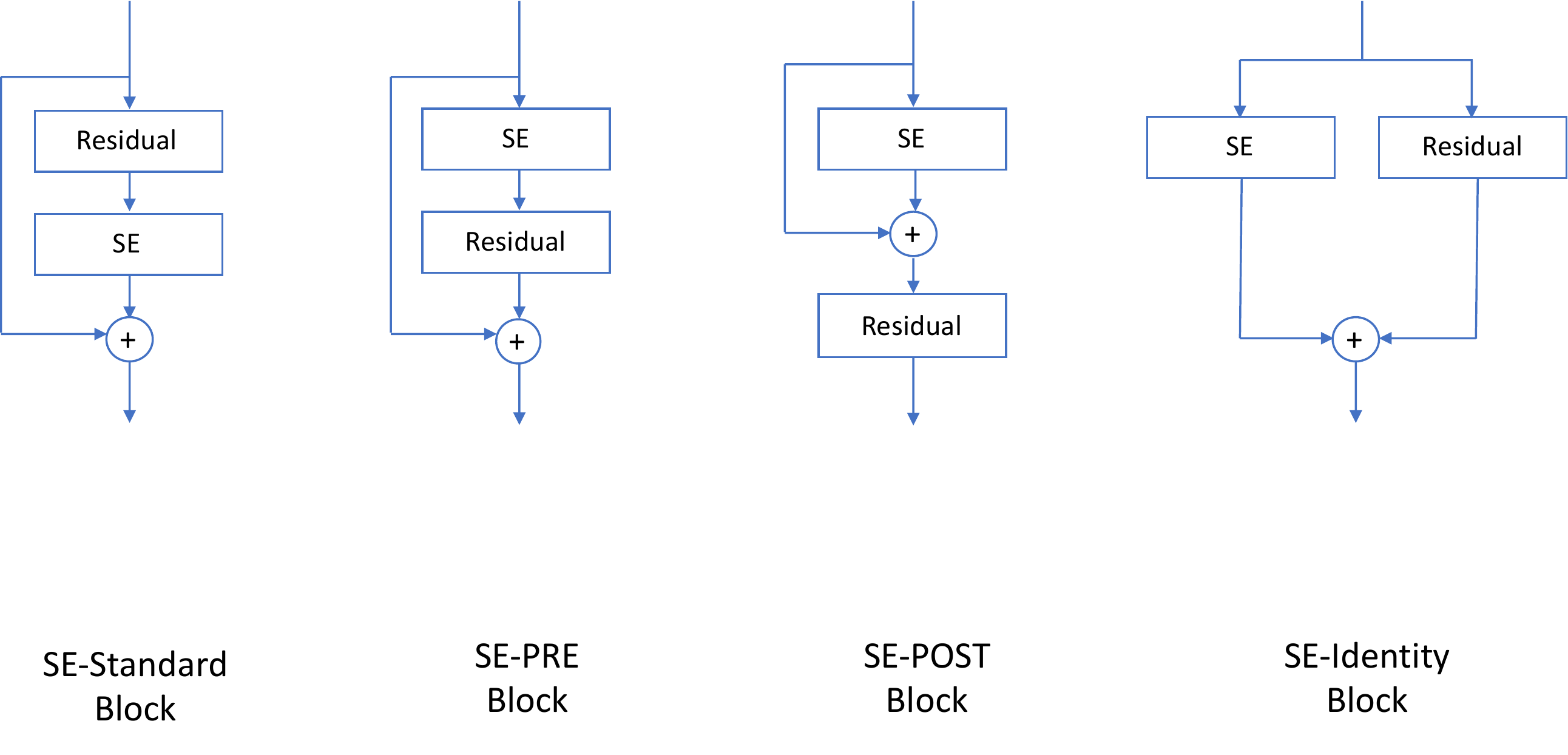}
\caption{Schema of different integration strategies of the SE blocks in ResNet architecture.}
\label{fig:integration}
\end{center}
\end{figure}

\begin{itemize}
\item \textbf{SE-PRE block} in which the SE block is moved before the residual unit.
\item \textbf{SE-POST block} in which the SE unit is moved after the summation with the identity branch (after ReLU).
\item \textbf{SE-Identity block} in which the SE unit is placed on the identity connection in parallel to the residual unit.
\end{itemize}

We observed that the SE-Standard design obtained the best performance in terms of EER or minDCF.

\begin{table}[H]
\center
\caption{Results obtained by using different integration strategies.}
\label{tbl:integration}
\resizebox{\columnwidth}{!}{
\begin{tabular}[|c]{|l|l|c|c|c|c|c|c|}
\hline
\textbf{System}  & \multicolumn{2}{c|}{\textbf{VoxCeleb1}} & \multicolumn{2}{c|}{\textbf{VoxCeleb1}} & \multicolumn{2}{c|}{\textbf{SITW}}\\
& \multicolumn{2}{c|}{\textit{-E cleaned}} & \multicolumn{2}{c|}{\textit{-H cleaned}} & \multicolumn{2}{c|}{\textit{core-core}}\\
 & EER & DCF & EER & DCF & EER & DCF \\\hline
SE-Standard  & \textbf{\eer{1.142}} & \dcf{0.1271} &  \textbf{\eer{1.985}} & \textbf{\dcf{0.1869}} & \textbf{\eer{1.312}} &  \textbf{\dcf{0.1105}}  \\
SE-PRE  & \eer{1.221} & \dcf{0.1316} &  \eer{2.091} & \dcf{0.1948} & \eer{1.476} &  \dcf{0.1174}  \\
SE-POST  & \eer{1.195} & \dcf{0.1337} &  \eer{2.041} & \dcf{0.1886} & \eer{1.367} &  \dcf{0.1209}  \\
SE-Identity & \eer{1.18} & \textbf{\dcf{0.1249}} &  \eer{2.034} & \dcf{0.1926} & \eer{1.367} &  \dcf{0.1131}  \\
\hline
\end{tabular}
}
\end{table}

\subsection{Different hidden layers}

Traditionally, the SE Blocks is composed of two hidden layers. Table~\ref{tbl:hiddenlayer} shows results when varying the number of hidden layers. The motivation behind that is to ensure that global information given by the pooling layer is well decorrelated by the different hidden layers. We observe that the SE block containing two hidden layers obtained the best results.

\begin{table}[H]
\center
\caption{Results obtained by systems using multi-statistical poolings.}
\label{tbl:hiddenlayer}
\resizebox{\columnwidth}{!}{
\begin{tabular}[|c]{|l|l|c|c|c|c|c|c|}
\hline
\textbf{System}  & \multicolumn{2}{c|}{\textbf{VoxCeleb1}} & \multicolumn{2}{c|}{\textbf{VoxCeleb1}} & \multicolumn{2}{c|}{\textbf{SITW}}\\
& \multicolumn{2}{c|}{\textit{-E cleaned}} & \multicolumn{2}{c|}{\textit{-H cleaned}} & \multicolumn{2}{c|}{\textit{core-core}}\\
 & EER & DCF & EER & DCF & EER & DCF \\\hline
h=1 & \eer{1.177} & \dcf{0.1294} &  \eer{2.04} & \dcf{0.1908} & \textbf{\eer{1.23}} &  \textbf{\dcf{0.1103}}  \\
h=2 & \textbf{\eer{1.142}} & \textbf{\dcf{0.1271}} &  \textbf{\eer{1.985}} & \dcf{0.1869} & \eer{1.312} &  \dcf{0.1105}  \\
h=3  & \eer{1.181} & \dcf{0.1274} &  \eer{2.056} & \dcf{0.1936} & \eer{1.312} &  \dcf{0.1125}  \\
h=4 & \eer{1.175} & \dcf{0.1296} &  \eer{2.013} & \textbf{\dcf{0.1858}} & \eer{1.367} &  \dcf{0.1134} 
 \\
\hline
\end{tabular}
}
\end{table}

\subsection{Pooling layer}

Table~\ref{tbl:pooling} investigates the performance by using different pooling layers in SE blocks. We propose to evaluate the performance of various poolings: 1) mean pooling (system called \emph{Mean}) and 2) maximum pooling (system called \emph{Max}). In addition to traditional poolings, we propose to evaluate: 1) standard-deviation pooling (system called \emph{Std}) and 2) the concatenation of mean- and standard-deviation poolings (system called \emph{Mean+Std}). We observe that the \emph{Mean+Std} system obtains the best performance.

\begin{table}[H]
\center
\caption{Results by using different pooling layers in SE Blocks.}
\label{tbl:pooling}
\resizebox{\columnwidth}{!}{
\begin{tabular}[|c]{|l|l|c|c|c|c|c|c|}
\hline
\textbf{System}  & \multicolumn{2}{c|}{\textbf{VoxCeleb1}} & \multicolumn{2}{c|}{\textbf{VoxCeleb1}} & \multicolumn{2}{c|}{\textbf{SITW}}\\
& \multicolumn{2}{c|}{\textit{-E cleaned}} & \multicolumn{2}{c|}{\textit{-H cleaned}} & \multicolumn{2}{c|}{\textit{core-core}}\\
 & EER & DCF & EER & DCF & EER & DCF \\\hline
Max  & \eer{1.21} & \dcf{0.1309} &  \eer{2.061} & \dcf{0.1934} & \eer{1.312} &  \dcf{0.1142}  \\
Mean & \eer{1.142} & \dcf{0.1271} &  \eer{1.985} & \textbf{\dcf{0.1869}} & \eer{1.312} &  \textbf{\dcf{0.1105}}  \\
Std & \eer{1.177} & \dcf{0.1294} &  \eer{1.994} & \dcf{0.1896} & \eer{1.312} &  \dcf{0.1130}  \\
Mean+Std & \textbf{\eer{1.134}} & \textbf{\dcf{0.1269}} &  \textbf{\eer{1.974}} & \dcf{0.1924} & \textbf{\eer{1.285}} &  \dcf{0.1132}  \\
\hline
\end{tabular}
}
\end{table}

\begin{figure*}
   \centering
    \subfigure[Stage 1]{\includegraphics[width=0.24\textwidth]{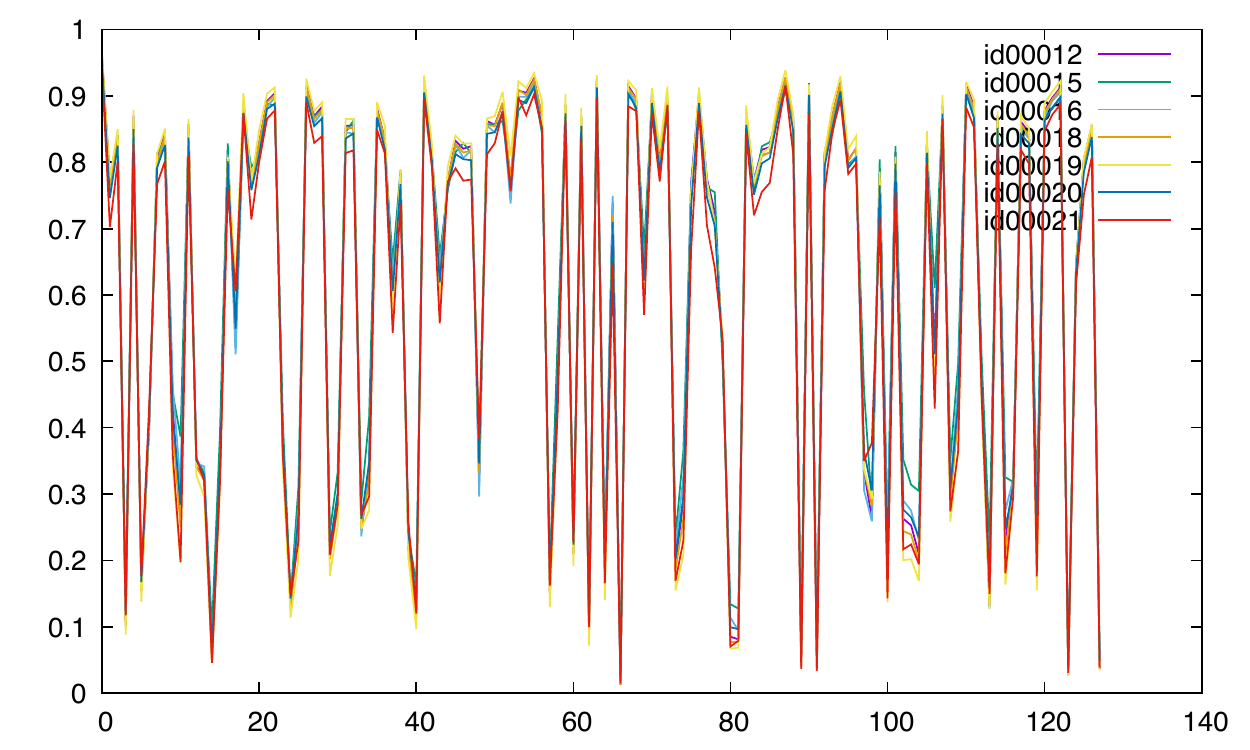}}
    \subfigure[Stage 2]{\includegraphics[width=0.24\textwidth]{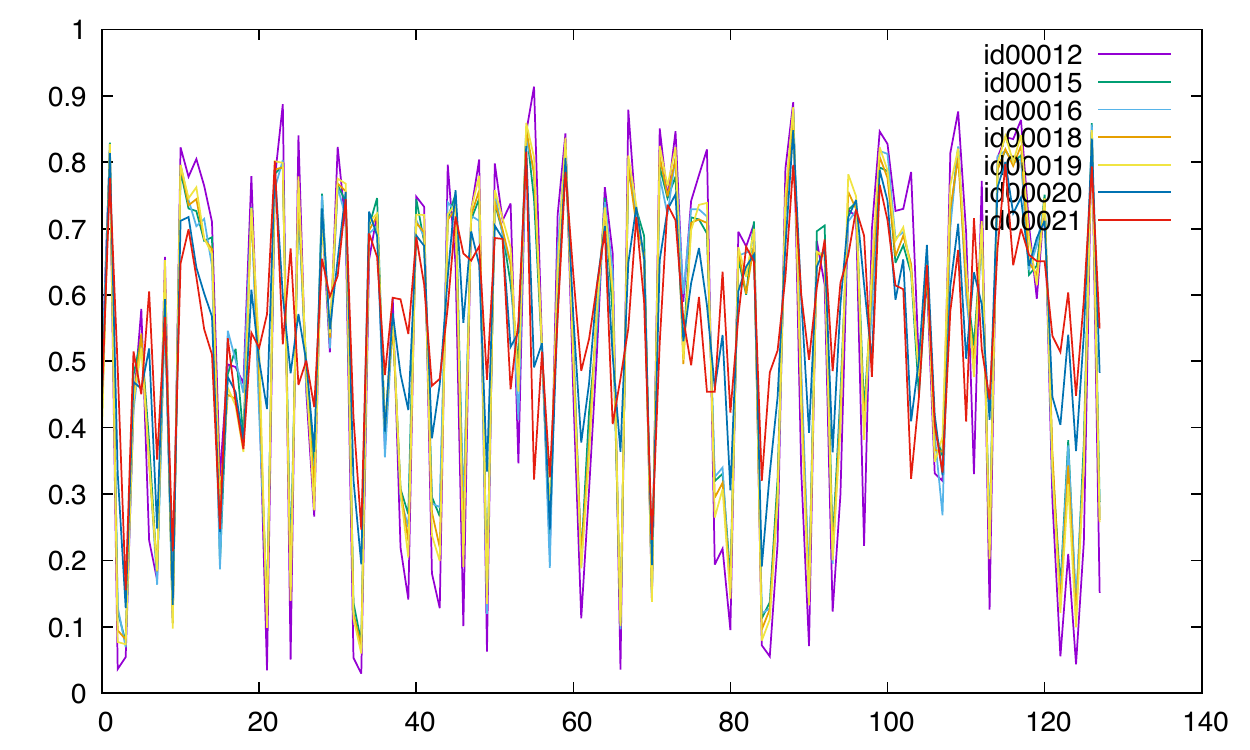}}
    \subfigure[Stage 3]{\includegraphics[width=0.24\textwidth]{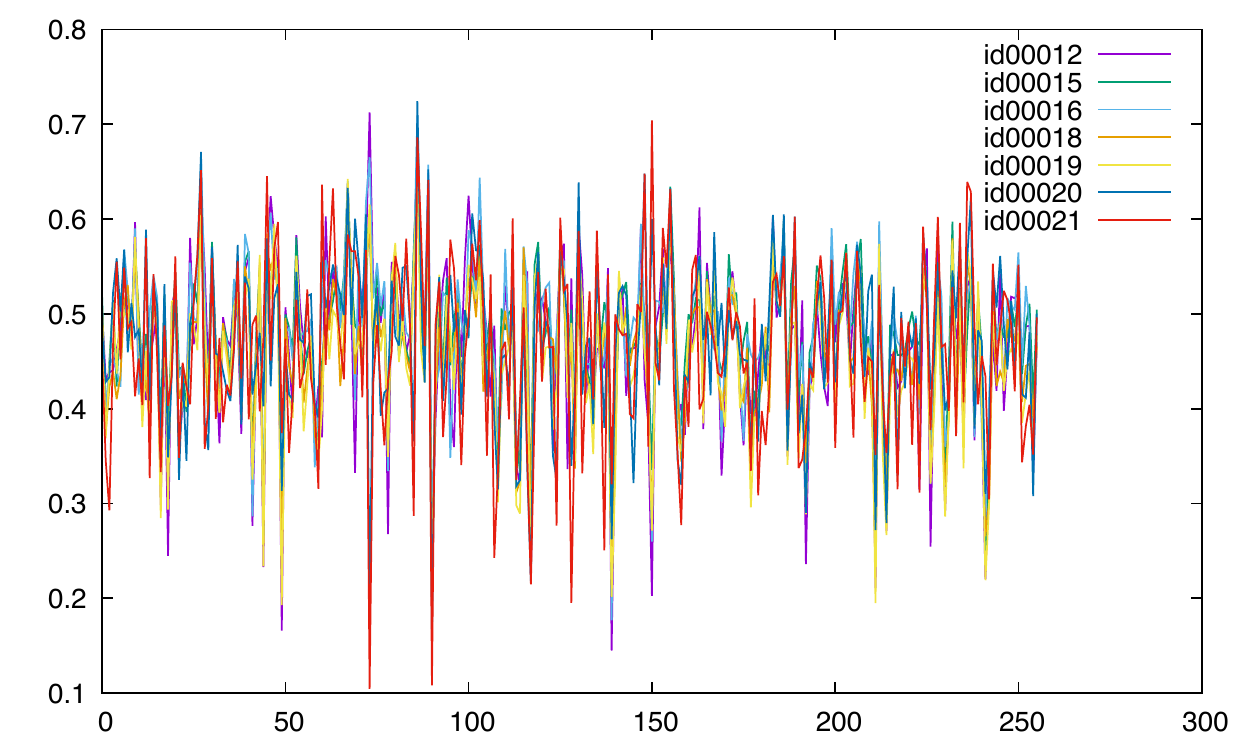}} \subfigure[Stage 4]{\includegraphics[width=0.24\textwidth]{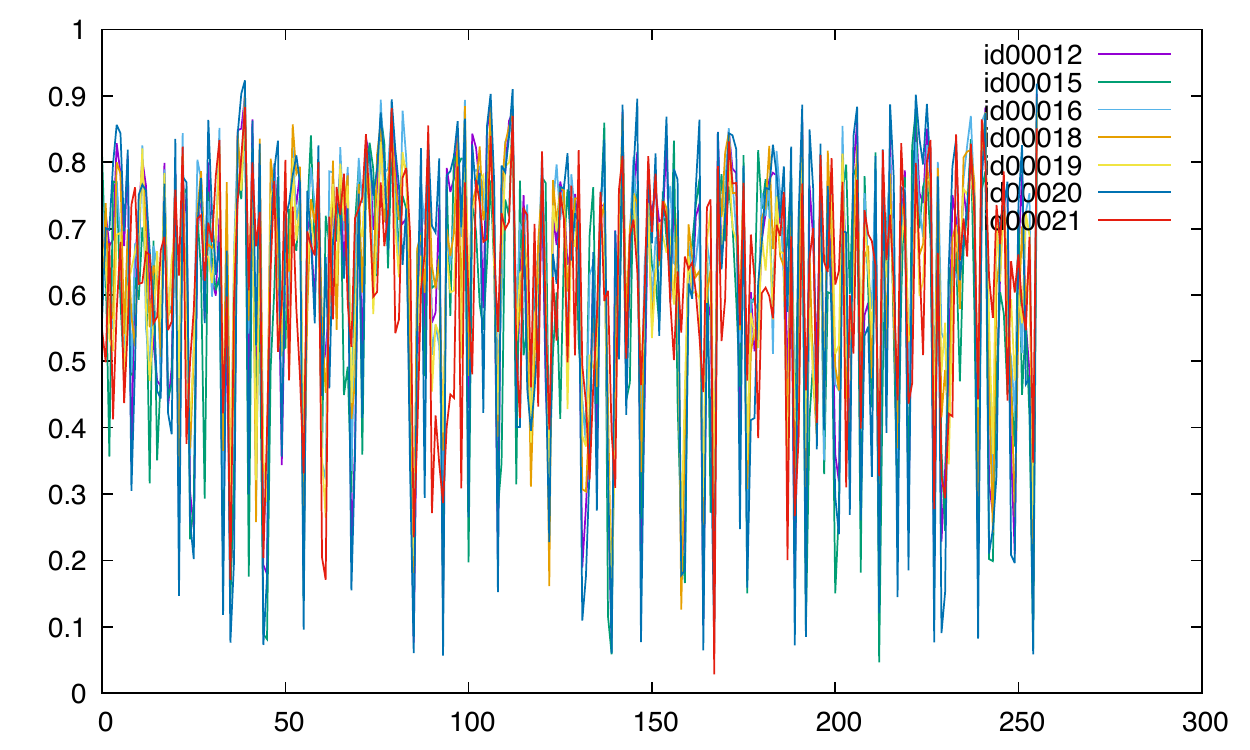}}
    \caption{Distribution of excitation across speakers given by the last SE blocks at different stages of ResNet.}
    \label{fig:multiple_speaker}
\end{figure*}

\begin{figure*}
   \centering
    \subfigure[Stage 1]{\includegraphics[width=0.24\textwidth]{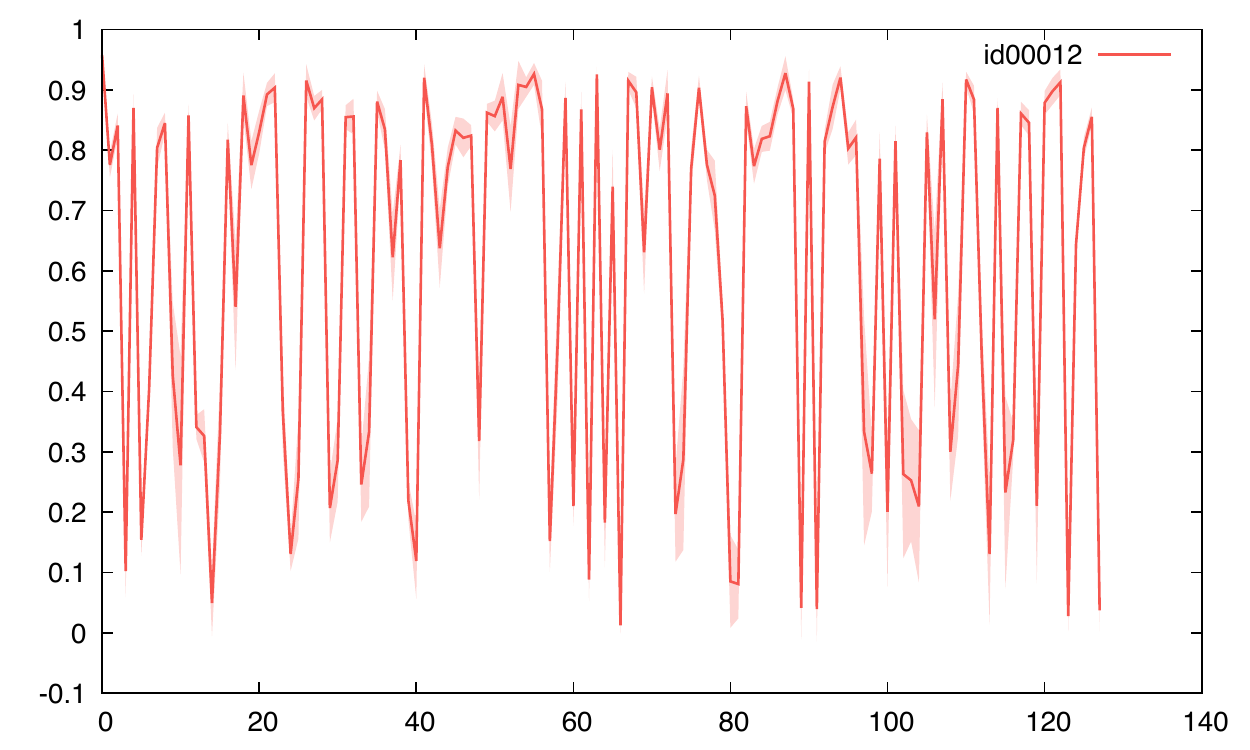}}
    \subfigure[Stage 2]{\includegraphics[width=0.24\textwidth]{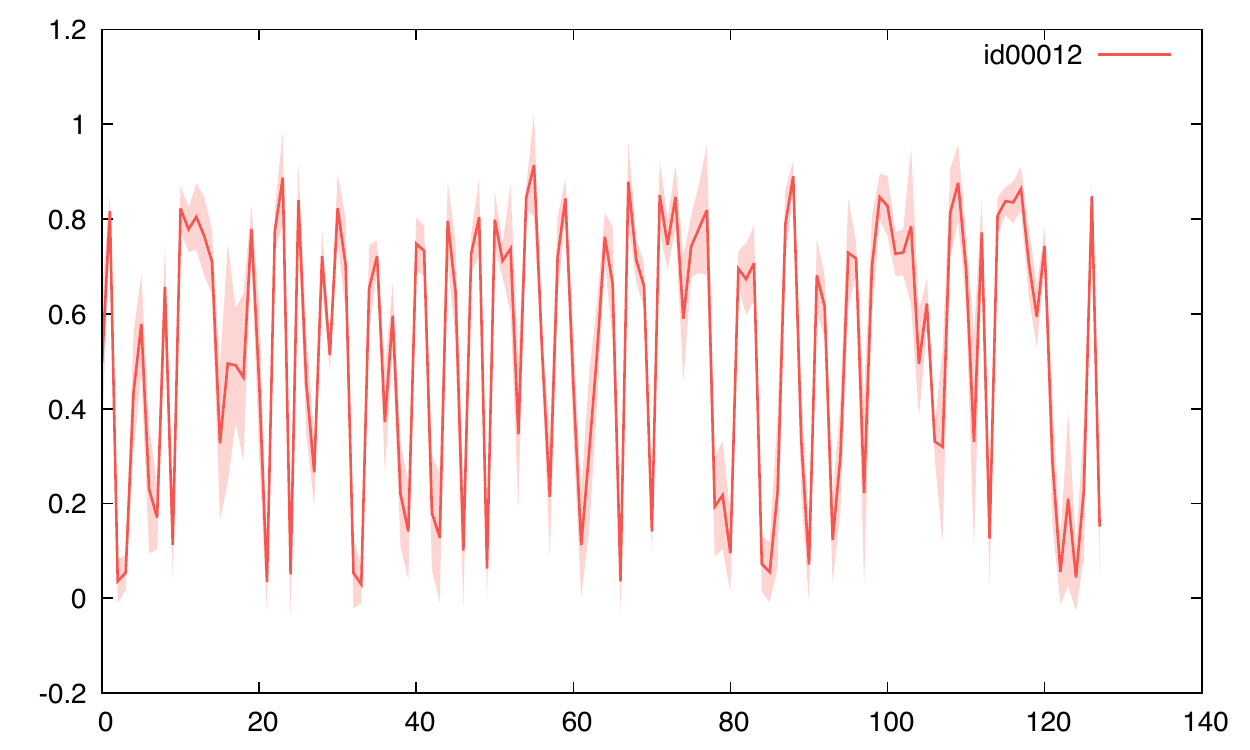}}
    \subfigure[Stage 3]{\includegraphics[width=0.24\textwidth]{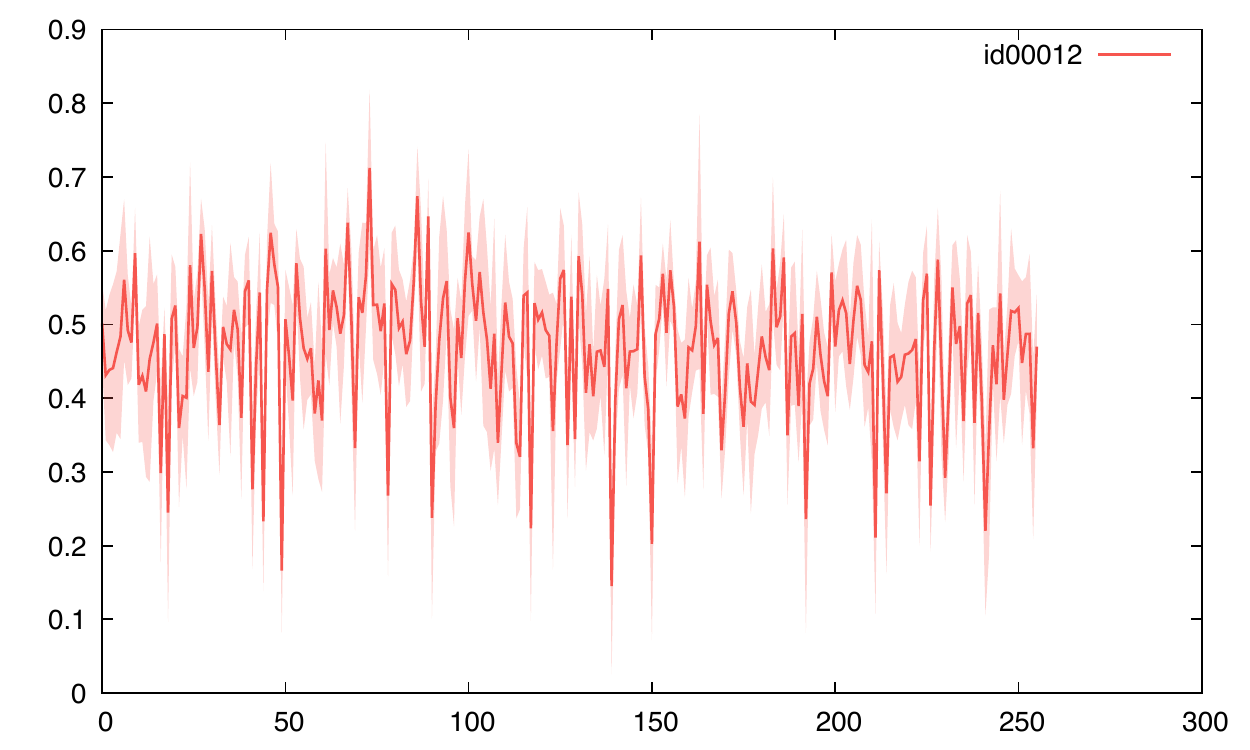}} \subfigure[Stage 4]{\includegraphics[width=0.24\textwidth]{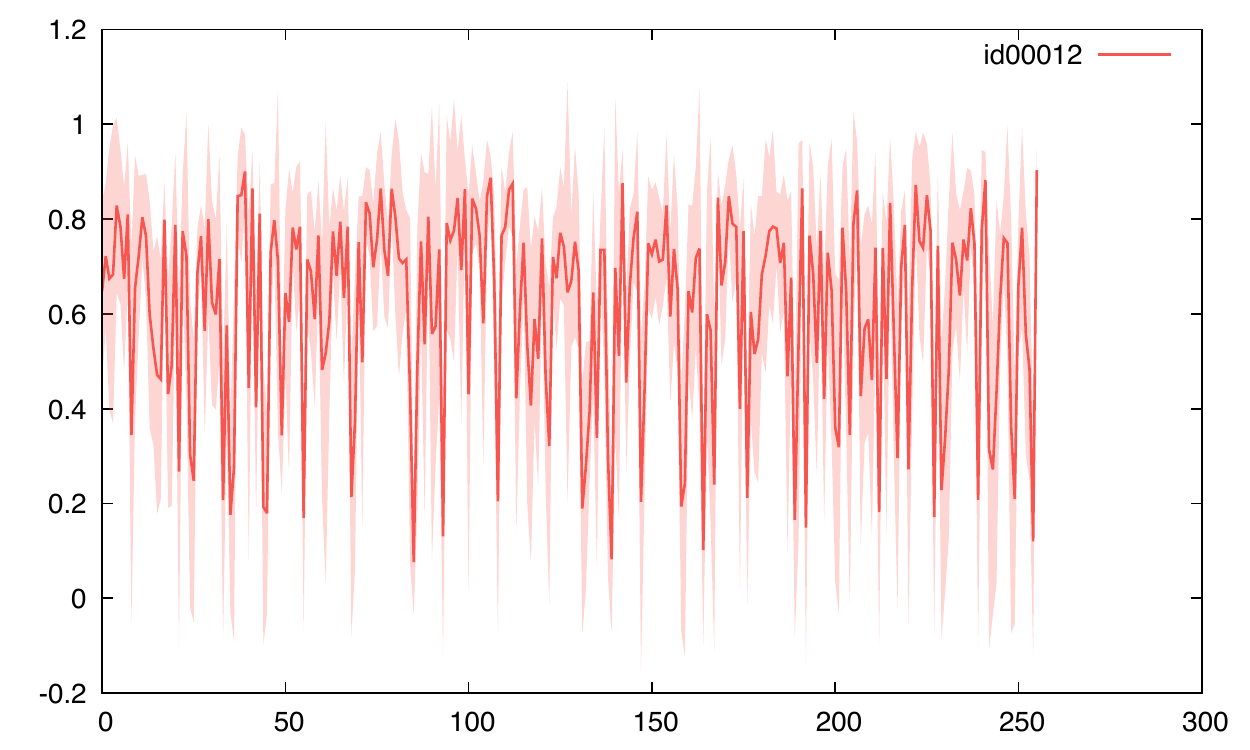}}
    \caption{Distribution of excitation within-speaker given by the last SE blocks at different stages of ResNet.}
    \label{fig:single_speaker}
\end{figure*}

\section{Role of Squeeze-and-Excitation}
\label{sec:study}

In this section, we study the role of squeeze-and-excitation in the context of speaker verification and, in particular, understand why SE is very efficient when is only integrated in Stage 1 and 2 of ResNet-34 architecture. We propose to study activations from the different SE blocks and their distribution at various stages in the network.

Fist, we study the distribution of excitation across speakers. Seven speakers are randomly picked up in the Voxceleb1 corpus. Then, for the last SE block of each stage, we compute, for each speaker, the mean activations of all segments. Figure~\ref{fig:multiple_speaker} depicts this distribution. It can be observed that the activation distribution is substantially the same at Stage 1, whatever the speaker (the lines of different speakers are overlap). However, the distribution is significantly varying from one speaker to another at Stage 4. We presume that the SE blocks used in low layers excite informative features in a class-agnostic manner, strengthening the shared high-level representations (Stages 1 and 2). In top layers, the SE blocks become increasingly specialised, and respond to different inputs in a highly class-specific manner (Stages 3 and 4).

Next, we study the within-speaker excitation distribution. Similarly to the previous experiments, we  randomly pick up one speaker in the Voxceleb1 corpus. Then, for the last SE block of each, we compute, for each speaker, the mean and standard activations of all the segments related to this speaker. Figure~\ref{fig:single_speaker} depicts this distribution. We observe that the standard deviation is rather weak at Stages 1 and 2, while it becomes more and more important at Stages 3 and 4. This reinforces our idea that low layers extract information independent to the speaker class while the high layers extract speaker-specific information.

\section{Conclusions}
\label{sec:conclusion}
In recent years, the introduction of the squeeze-and-excitation (SE) method has allowed to overcome some weaknesses of CNN architectures in the research field of image recognition. Since then, introduced in speaker verification, this method required to be adapted to the specificity of this research field.

In this paper, different architectures and configurations of SE are presented and evaluated in order to build a robust $x$-vector extractor for speaker verification. The results of our experiments show that SE blocks used in low-layers excite informative features in a class-agnostic manner. Moreover, when used in top layers, the SE blocks become increasingly specialised to class-specific manner.

Experiments performed on the SITW, Voxceleb1-E Cleaned and Voxceleb1-H dataset showed significant gains by using SE blocks at Stage 1 and 2 and by using a pooling layer combining mean- and standard-deviation statistics (leading to a relative gain of 9\% in terms of equal error rate). These experiences confirm the need to properly adapt the architecture and configuration of SE to the task of speaker verification.

\section{\textbf{Acknowledgement}}
\label{s:acknowledge}
This research was supported by the ANR agency (Agence Nationale de la Recherche), RoboVox project (ANR-18-CE33-0014).

\bibliographystyle{IEEEbib}

\bibliography{mybib}

\end{document}